\documentclass[preprint2,apj,amsmath,amssynb]{aastex}

\newcommand{\pubjournal}[4]{#4, #1, #2, #3}

\shorttitle{Inverse Compton Gamma-rays from the Galactic Centre}

\shortauthors{Hinton \& Aharonian}

\begin{document}
  
  \title{Inverse Compton Scenarios for the TeV Gamma-Ray Emission 
    of the Galactic Centre
  }
  

\author{J. A. Hinton} 

\affil{Max-Planck-Institut f\"ur Kernphysik, P.O. Box 103980, D 69029 Heidelberg, Germany}
\affil{Landessternwarte, Universit\"at Heidelberg, K\"onigstuhl, D 69117 Heidelberg, Germany}
\affil{School of Physics \& Astronomy, University of Leeds, Leeds LS2 9JT, UK}

\author{F. A. Aharonian} 

\affil{Dublin Institute for Advanced Studies, 5 Merrion Square, Dublin 2, Ireland}
\affil{Max-Planck-Institut f\"ur Kernphysik, P.O. Box 103980, D 69029 Heidelberg, Germany}



\begin{abstract}

The intense Compton cooling of ultra-relativistic electrons in the 
Klein-Nishina regime in radiation dominated 
environments, such as that found in the Galactic Centre, may result in
radically different electron spectra than those produced by
Synchrotron cooling.  We explore these effects and their impact on the
X-ray and $\gamma$-ray spectra produced in electron accelerators in
this region in comparison to elsewhere in our galaxy. We discuss the
broad-band emission expected from the newly discovered pulsar wind
nebula G\,359.95$-$0.04 and the possible relationship of this X-ray
source to the central TeV $\gamma$-ray source HESS\,J1745$-$290.
Finally we discuss the possible relationship of the Galactic Centre 
INTEGRAL source IGR\,J1745.6$-$2901 to the TeV emission.

\end{abstract}

\keywords{pulsar wind nebulae: general --- pulsar wind nebulae:
  individual(G\,359.95$-$0.04, G\,0.9+0.1)}

\section{Introduction}

The detection of TeV $\gamma$-rays from the Galactic Centre (GC) by
several groups, \cite{whipple,cangaroo,hess_gc2003, magic_saga}, can be
considered as one of the most exciting discoveries of recent years in
High Energy Astrophysics. After initial disagreements, the basic
properties of the TeV source (HESS\,J1745$-$290) now seem to be firmly
established, with the values from the H.E.S.S.~\citep{hess} instrument
providing the highest level of accuracy. The key experimental findings
are:

\begin{itemize}

\item The energy spectrum in the range 0.15--20 TeV can be described by a power
law: $dN/dE = k (E/\mathrm{1 TeV})^{-\Gamma}$ cm$^{-2}$ s$^{-1}$ TeV$^{-1}$
with $k=1.8\pm0.1_{stat}\pm0.3_{sys}, \Gamma=2.29\pm0.05_{stat}\pm0.1_{sys}$ 
\citep{hess_gc2004_icrc}.

\item There is no evidence for variability on hour to year time-scales~\citep{hess_gc2004_icrc, magic_saga}

\item The centroid of the $\gamma$-ray emission lies within $1'$ of Sgr~A$^{\star}$~\citep{hess_gc2003, hess_gc2004_icrc}

\item The rms size of the emission region must be less than $3'$, equivalent to $7$ parsecs
at the GC distance~\citep{hess_gc2003}

\end{itemize}

The implied 1--10 TeV $\gamma$-ray luminosity of the source is
$10^{35}$ erg s$^{-1}$.  A wide range of possible counterparts and
mechanisms have been put forward to explain the $\gamma$-ray
emission. These include the annihilation of dark matter (unlikely due
to the measured spectral shape, see~\citet{profumo}) and the
astrophysical objects Sgr~A$^{\star}$ \citep{aharonian_neronov} and Sgr~A~East
\citep{fatuzzo,crocker}. A hadronic
origin of the $\gamma$-ray emission seems plausible, originating
either within these sources or indirectly via the injection of hadrons
into the dense central parsec region~\citep{aharonian_neronov,lu,liu}.
Indeed, there is strong evidence for the existence of a proton
accelerator close to the GC (at least in the past) in the form of the
recently discovered TeV emission of the GMCs of the central molecular
zone~\citep{hess_gcdiffuse}. Nevertheless, an origin of the central
$\leq 0.1^{\circ}$ $\gamma$-ray emission in the interactions of TeV electrons 
remains a compelling alternative. Several scenarios have
been discussed in which the persistent TeV emission is explained by
inverse Compton (IC) scattering of electrons in the central
parsec. The termination shock of a hypothetical wind from the
supermassive black hole~\citep{atoyan_dermer}, stellar wind
shocks~\citep{loeb} and the newly discovered X-ray nebulae $8''$ from
Sgr~A$^{\star}$~\citep{wang} have all been proposed as acceleration
sites for these electrons.

\begin{figure}
\epsscale{1.05}
\plotone{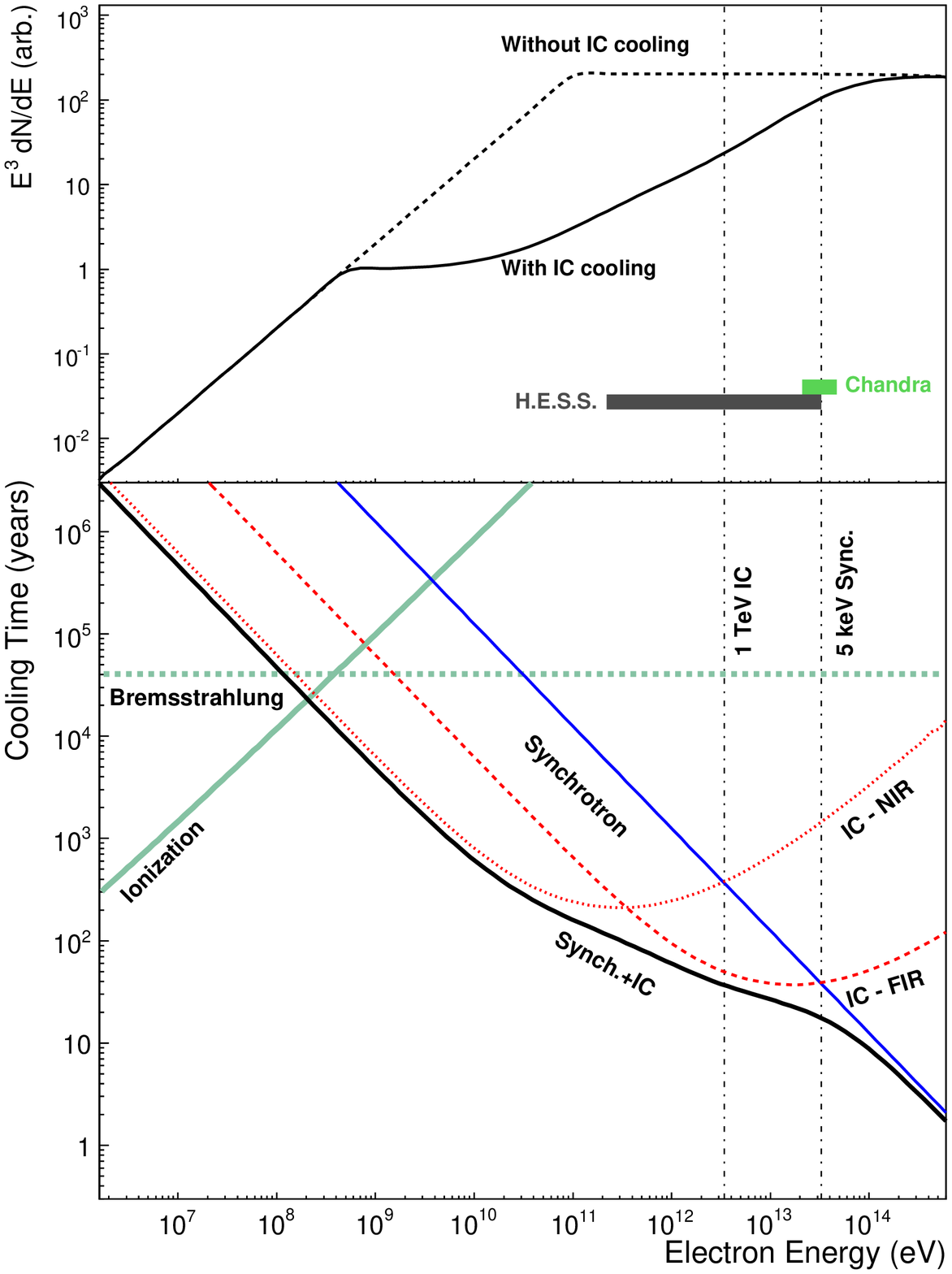}
\caption{
  Top panel: energy spectrum of electrons with continuous
  injection with $dN/dE \propto E^{-\alpha}$ (with $\alpha=2$) 
  and cooling over a
  $10^{4}$ year period. The dashed line shows the cooled spectrum
  for electrons suffering only synchrotron losses (for $B = 100$ $\mu$G). 
  The solid line shows the spectrum after synchrotron and IC cooling
  on radiation fields typical of the central parsec of our galaxy.
  The shaded regions show the range of electron energies contributing
  to to signals seen in the energy ranges of the Chandra and H.E.S.S.
  instruments. Bottom panel: cooling time via IC (dashed lines, FIR and optical
  radiation fields) and synchrotron radiation (solid blue line).
  The lower heavy line shows the overall cooling time for IC and 
  synchrotron radiation. The approximate energy loss time-scales
  for ionisation and bremsstrahlung (in a neutral environment of number 
  density 1000 cm$^{-3}$) are shown for comparison.
}
\label{f1}
\end{figure}

Whilst the formation of synchrotron and IC nebulae around sources of
multi-TeV electrons proceeds in general in the GC as in other regions
of the galactic disk (GD), the very high density of low-frequency
radiation in GC leads to significant deviations from the typical disk
scenario.  The high radiation density (out to $\sim10$ pc from
Sgr~A$^{\star}$) not only provides copious targets for $\gamma$-ray
production, but also creates rather unusual conditions for the
formation of the spectrum of TeV electrons. For magnetic fields less
than $\sim$ 100 $\mu$G, the energy density of radiation appears much
higher than the energy density of the magnetic field, thus even in the
modest Klein-Nishina (KN) regime, TeV electrons are cooled
predominantly by IC losses. This leads to \emph{hardening} of the
spectrum (not steepening as in the typical GD environment) up to very high
($\geq 100$ TeV) energies (the deep KN regime), where the synchrotron
losses start to dominate over IC losses.  
Figure~\ref{f1} illustrates the cooling time for electrons in
the presence of both strong radiation fields and magnetic fields 
(lower panel) and the modification of the injected electron 
spectrum after cooling (upper panel).
The time-evolution of the electron spectrum (in Figure~\ref{f1} and throughout 
this paper) is calculated numerically, considering energy losses and
injection of electrons in time-steps much
shorter than the age of the system. Synchrotron and IC energy losses 
are calculated using the formalism developed by \citet{blumenthal}.

\begin{figure}
\plotone{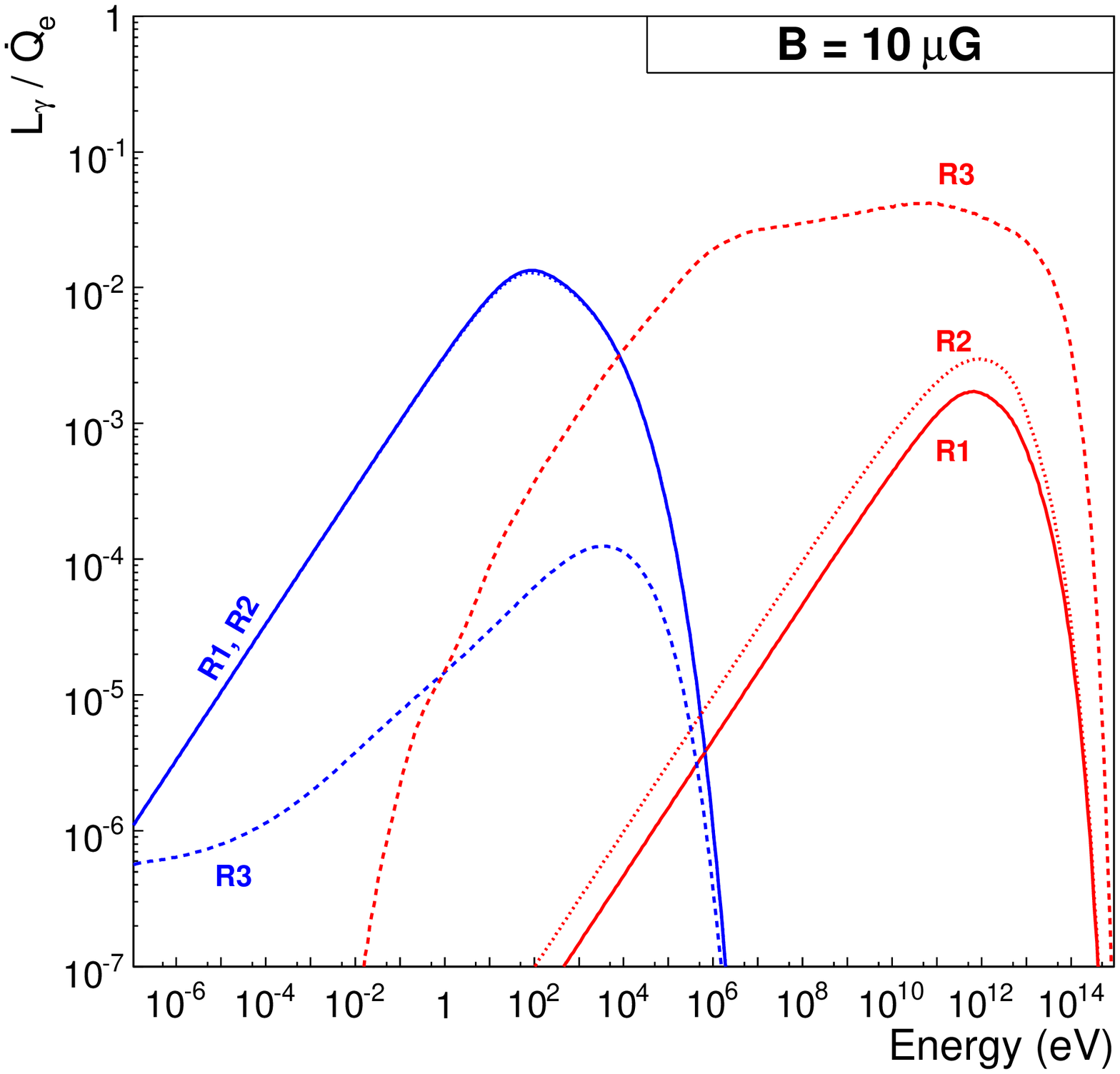}
\plotone{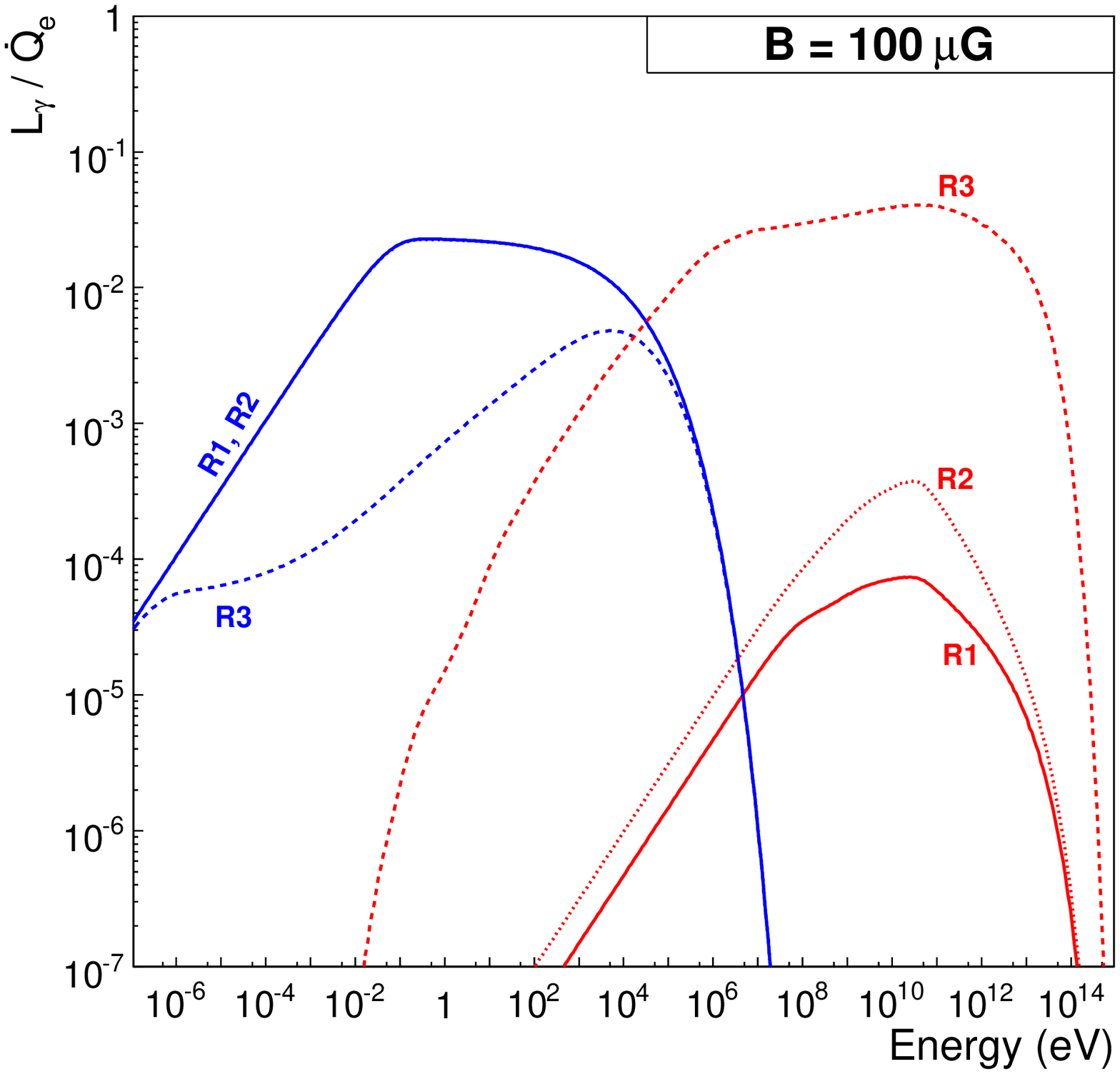}
\caption{ Fraction of the total power injected in electrons radiated
  in different spectral bands for $B=10$~$\mu$G (top) and $B=100$~$\mu$G
  (bottom).  In each case three radiation fields are considered: R1,
  a typical galactic disk environment; R2, $\sim$100~pc from the
  GC and R3, within the central parsec.  Continuous injection 
  (over a $10^{4}$ year period) with $dN/dE \propto
  E^{-2}$ and an exponential cut-off at 100~TeV 
  are assumed in all cases. The synchrotron
  curve for R2 lies underneath that for R1.
}
\label{f2}
\end{figure}

The irregular spectral shape of electrons shown in Fig.~\ref{f1} 
is reflected differently in the synchrotron and IC radiation
components \citep[see e.g. ][]{khangulyan,moderski}.
Another interesting feature of these conditions is that due to enhanced IC
losses, the synchrotron radiation of electrons will be strongly
suppressed (by an order of magnitude or more), unless the magnetic
field in extended regions of GC exceeds 100 $\mu$G.
This situation is in stark contrast to that in the GD.
where IC emission is strongly suppressed for magnetic fields $>10$ $\mu$G.
In Fig.~\ref{f2}, we compare the fractional energy distribution 
resulting from the injection of the same power-law spectrum of
electrons for two values of the mean magnetic field in the source
(10 $\mu$G and 100 $\mu$G) for 3 locations in our galaxy: (i) 
the central 1~pc, (ii) at 100~pc from GC, and (iii) in a standard site in the GD.  
The radiation fields used in Fig.~\ref{f2} and throughout this
paper are given in Table~\ref{t1}.

\begin{table*}
\begin{center}
\begin{tabular}{|l|c|c|c|c|}\tableline
Rad. Field / Component ($kT$)  & UV/optical & NIR & FIR & CMBR\\ 
 & (3 eV) & (0.3 eV) & ($6\times10^{-3}$ eV) & ($2.35\times10^{-4}$ eV) \\ \tableline
R1 & - & 0.2 & 0.2 & 0.26 \\
R2 & - &  9   & 1   & 0.26 \\
R3 & 5000 & 5000 & 500 & 0.26 \\
R4 & 50 & 50 &  5 &  0.26 \\\tableline
\end{tabular}
\end{center}
\caption{ Energy density (in eV cm$^{-3}$) of the thermal components
  of radiation fields used for the calculation of electron
  cooling and $\gamma$-ray production via inverse Compton
  scattering. Field R1 represents the typical galactic disk
  environment. R2 reflects the situation in the inner $\sim$100 pc and
  R3 is a model of the intense field of the central cubic parsec of
  our galaxy.  R4 is a scaled down version of R3 to approximate the
  situation 10 pc from the GC.
  \label{t1}
} 
\end{table*}

Some of these effects have been discussed by \citet{wang} in the
context of a possible identification of HESS\,J1745$-$290 with the
candidate Pulsar Wind Nebula (PWN) G\,359.95$-$0.04.  In this
paper, we present the results of numerical calculations based on
time-dependent treatment of the formation of the energy spectrum of
electrons.  We discuss the case of G\,359.95$-$0.04 and show that indeed
this PWN can explain the TeV $\gamma$-ray emission from the GC. The
implied $B$-field in this scenario appears to be around 100 $\mu$G.
Remarkably, PWNe similar to G\,359.95$-$0.04 (i.e. with similarly large
$B$-field and comparable energetics) located in the conventional sites
within the GD would be undetectable with any current or planned TeV
$\gamma$-ray instrument. The converse also holds: typical TeV
$\gamma$-ray PWNe (with $B$-field of about 10 $\mu$G or less) would not
be detectable in X-rays if located in the central 1~pc region.
Finally, we discuss the conditions implied by the interpretation of the
hard X-ray emission detected by INTEGRAL~\citep{neronov,belanger} as
synchrotron emission of multi-TeV electrons in the context of the
severe IC losses of these electrons.

\section{Pulsar Wind Nebulae}

Pulsar Wind Nebulae are perhaps the most efficient astrophysical
particle accelerators in our galaxy. The best studied PWN, the Crab
Nebula, accelerates electrons up to $\sim10^{16}$~eV despite the rapid
synchrotron losses of these particles in its $160$~$\mu$G magnetic
field~\citep{hegra_crab}. The recent detections of extended TeV
emission from several PWNe, including MSH-15$-5$-{\it
02}~\citep{hess_msh1552} and
G\,18.0-0.7/HESS\,J1825-137~\citep{hess_j1825} with the H.E.S.S.
instrument suggest that such objects are copious TeV $\gamma$-ray
emitters. In this context, a PWN may provide a natural explanation for
the GC TeV emission. Here we discuss, in detail, the case of the new PWN
candidate G\,359.95-0.04.

\subsection{The case of G\,359.95-0.04}

The X-ray nebula G\,359.95-0.04 was discovered in deep Chandra
observations of the Galactic Centre~\citep{wang} and lies at a
projected distance to Sgr~A$^{\star}$ of 0.3~pc. The nebula exhibits a
cometary morphology with a projected size of 0.07 $\times$ 0.3
pc. The overall energy spectrum of this object is purely non-thermal,
with a power-law index of $1.94^{+0.17}_{-0.14}$ and an unabsorbed
2--10 keV X-ray luminosity of $\approx10^{34}$ erg/s.  The Chandra
data reveal a softening of spectral index with distance from the
``head'' of the nebula, a possible signature of cooling of electrons
away from the accelerator. \citet{wang} have suggested that the head
of the nebula contains a young pulsar and that G\,359.95-0.04 is
likely a ram-pressure-confined PWN.

G\,359.95-0.04 lies within the 68\% confidence error circle of the
$\gamma$-ray source HESS\,J1745-290.  The possible connection
between these two objects was pointed out by \citet{wang}, who discuss
in some detail the important physical aspects involved in the
relationship between the X-ray and $\gamma$-ray emission. One of our
aims here is to make a full time-dependent calculation to investigate
more deeply the likelihood of an association of these two objects.  A
major difficulty in such an association is the $\sim0.1^{\circ}$ angular resolution
of H.E.S.S. In this scenario, the
$\gamma$-ray signal would be point-like and non-variable and the only
available information for modelling is the spectral data in the X-ray
and $\gamma$-ray bands. However, the X-ray morphology provides some
clues to the environment of the PWN. For example, the fact that X-ray
spectrum softens rather than hardens away from the (nominal) pulsar
position indicates that the X-ray emitting electrons are cooled by
synchrotron radiation (or IC radiation in the Thompson regime) rather
than by IC radiation in the Klein-Nishina (KN) regime as might be
expected in the dense GC radiation fields. This fact alone places a lower
limit on the magnetic field in the PWN of $\sim100$ $\mu$G.

\begin{figure}[ht]
\plotone{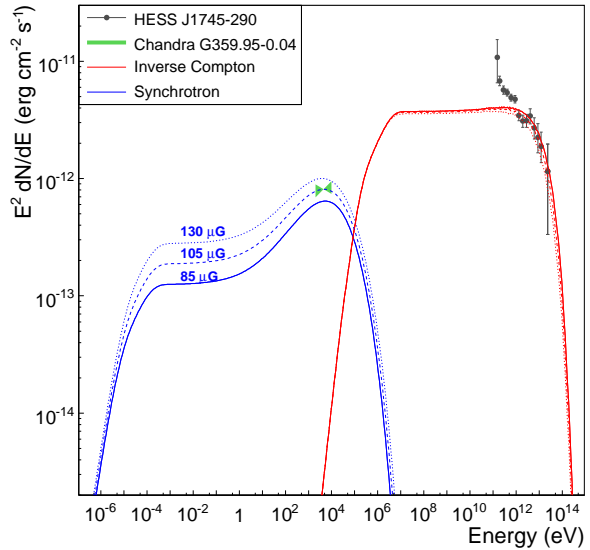}
\caption{
  Spectral energy distribution for inverse Compton scattering on
  a single temperature FIR radiation field of density 5000 eV\,cm$^{-3}$.
  The three line styles indicate the effect of changing magnetic
  field strength (with all other parameters fixed). The assumed
  injection spectrum is described in the main text. H.E.S.S. data 
  are taken from \citep{hess_gc2004_icrc}, Chandra data from \citep{wang}.
}
\label{f3}
\end{figure}

Due to KN suppression, it is likely that the dominant target for IC
radiation at a few TeV is the far infrared background. Matching the
flux of HESS\,J1745$-$290 at these energies with a nominal FIR
radiation energy density of 5000~eV~cm$^{-3}$ \citep{davidson}
requires a $B$-field of $\approx\,105$~$\mu$G.  In the case that HESS\,J1745$-$290
and G\,359.95-0.04 are \emph{not} associated, this value provides a
lower limit on the average magnetic field in the PWN.  Fig.~\ref{f3}
shows a model spectral energy distribution (SED) for G\,359.95$-$0.04
with IC on a FIR field.  The injection spectrum of electrons is
assumed to begin at 1 GeV and be of the form $dN/dE \propto E^{-\alpha} e^{-E/E_{0}}$
with $\alpha$ = 2 and $E_{0} = 100$ TeV. A source age of $10^{4}$
years is assumed and a total power of $6.7 \times 10^{35}$ erg/s
injected into relativistic electrons is required to match the measured
X-ray flux (assuming a distance to the galactic centre of $7.6\pm0.4$
pc \citep{gc_distance}).  As the cooling time of the electrons
responsible for the observed X-ray and $\gamma$-ray emission is much
shorter than the age of the pulsar in this scenario, possible evolutionary
effects on the injection power (related to the breaking of the pulsar spin) 
can safely be neglected, we therefore
assume a constant injection rate in the simulations presented here.
Fig.~\ref{f3} demonstrates an important aspect of IC cooling: the KN
effect acts twice on the IC spectrum (firstly by distortion of the
electron spectrum through cooling, and secondly in the production of
IC emission), but only once on the Synchrotron spectrum. This means
that the hardening effect of cooling in the KN regime is masked in the
IC spectrum but clearly visible in the Synchrotron emission.

\begin{figure}[h]
\plotone{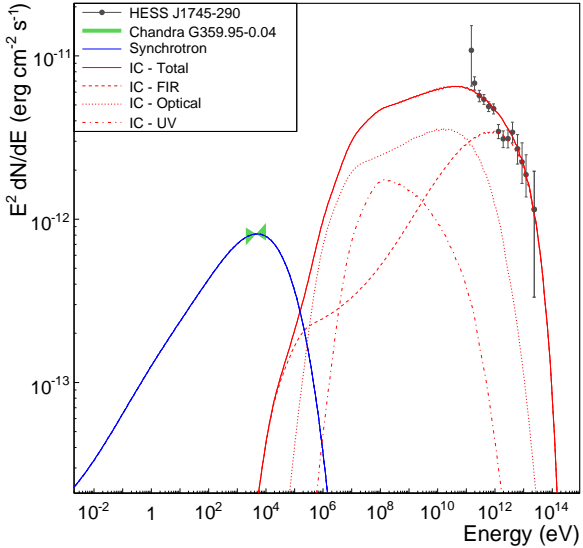}
\caption{
  Spectral energy distribution for a realistic GC radiation field 
  with FIR, optical and UV components. The magnetic field strength
  is fixed at 105~$\mu$G. Details are given in the main text.
  \label{f4}
}
\end{figure}

The effect of adding different temperature components to the GC
radiation field is shown in Fig.~\ref{f4}. Optical ($kT = 0.3$ eV),
and ultraviolet ($kT = 3.0$ eV) energy densities of $5\times10^{4}$
eV\,cm$^{-3}$ are assumed, consistent with the values expected within the
central parsec of our galaxy \citep{davidson}. The injected electron
spectrum is identical to that in Fig.~\ref{f3}.  On such a compound
field, low energy electrons are cooled by IC scattering on optical seed photons,
with higher energies cooled by IC on the FIR. This effect leads to
rather different shapes for the IC spectra from these two
components. It can be seen in this figure that the contribution of UV is likely
to be small because of strong KN suppression. This optical/UV domain
can in principle be explored by the GLAST satellite \citep{glast}, but
such measurements may be rather difficult due to the strong diffuse
background and the modest angular resolution of the instrument.
As is clear from Fig.~\ref{f1}, Bremsstrahlung
losses are unlikely to be important in the PWN as the ambient density
is likely $\ll 1000$ cm$^{-3}$.

\begin{figure}[h]
\plotone{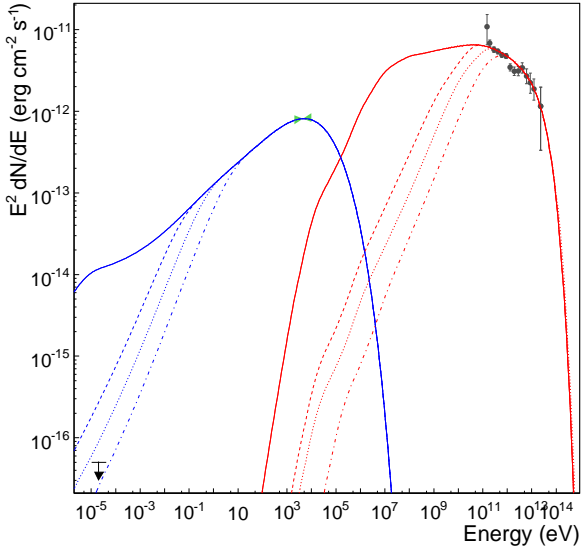}
\caption{
  As for Fig~\ref{f4} but showing the effect of a low energy cut-off
  in the injected electron spectrum (and with an expanded flux scale). 
  The corresponding minimum energies
  are 1 MeV (solid line), 100 GeV (dashed line), 300 GeV (dotted line) and 
 1 TeV (dot-dashed line). A 6cm radio flux upper limit for
G\,359.95-0.04 from \citet{yusefzadeh_pc} is shown.
  \label{f4b}
}
\end{figure}

The spectral and spatial distribution of low energy electrons in the
PWN can in principle be traced using radio observations. 
However, no point-like or extended source is observed at
the position of G\,359.95-0.04 in 6~cm observations and
a three sigma energy flux upper limit of $5 \times 10^{-17}$  erg cm$^{-2}$ s$^{-1}$
has been derived~\citep{yusefzadeh_pc}.
This limit lies almost three orders of
magnitude below the curve shown in Fig.~\ref{f3}. There are two 
factors which may both act to mitigate this apparent contradiction: 

\begin{enumerate}
\item the electron energy spectrum has a low energy cut-off.
Figure~\ref{f4b} shows the impact of a low energy cut-off in the electron
spectrum on the IC and synchrotron spectra. If the radio emission 
region is considered to be identical to that of keV X-rays (see point 2
below) then there is an implied low energy cut-off at $\sim$ 1~TeV
and rather poor agreement with the lowest energy $\gamma$-ray data points.
Indeed a low energy cut-off in the electron spectrum is expected within the 
PWN paradigm. For example, in the case of the Crab Nebula, \citet{kennel}
suggest a minimum injection energy of $\sim$ 1 TeV at the wind termination 
shock. \citet{wang} suggested a minimum low energy cut-off of at 
5~GeV derived from the radio data available at that time.

\item the cooling time of the radio emitting electrons is almost three
orders of magnitude longer than that of the X-ray emitting
electrons. Depending on the transport mechanism of particles in the
nebula the angular diameter of the radio emission may be expected to
be much larger than the X-ray nebula and with a correspondingly lower
surface brightness. We envisage two general transport scenarios:
a) \emph{energy independent advection}. In this case the PWN size is 
    inversely proportional to the cooling time of electrons and
    the ratio of radio to X-ray angular size is $\sim$1000 and
    the flux within the bounds of the X-ray nebula is a factor of $10^{6}$
    lower than that presented in figures \ref{f4} and \ref{f4b}.
    Clearly there is no contradiction to the radio limit in this case. 
b) \emph{diffusion}  with  $D \propto E^{\alpha}$ and $r = \sqrt{2Dt}$.
   In this case the PWN size is proportional to $\sqrt{E^{\alpha}/t_{\mathrm{cool}}}$.
   For a value of $\alpha=0.5$ the expected ratio of radio to X-ray
   size is $\sim 3$ (implying a one order 
      of magnitude reduction in surface brightness). In this case a low
      energy cut-off at $\sim50$ GeV is still required.
   Larger values of $\alpha$ (for example Bohm diffusion, $\alpha=1$) 
   appear to be excluded by the observed energy dependant 
      morphology at X-ray wavelengths.
\end{enumerate}
Regardless of the nature of the TeV source, 
it seems that one or both of these effects must occur to explain 
the radio to X-ray behaviour of G\,359.95-0.04. 

Fig.~\ref{f1} (top panel) illustrates the electron energies
contributing to the H.E.S.S. and Chandra signals for a 100 $\mu$G
magnetic field. As the $\gamma$-ray emission takes place predominantly in
the KN regime, the energy range of electrons probed by H.E.S.S. is
extended by an order of magnitude relative to the Thompson regime
case. In contrast, the narrow energy range probed by Chandra reflects
the standard $\epsilon_{\gamma} \propto \sqrt{\epsilon_{e}}$ case.  As
synchrotron emission below 1~keV is heavily absorbed in the GC, VHE
$\gamma$-ray emission represents the \emph{only} way to study the
200~GeV -- 20~TeV electrons. From the lower panel of Fig.~\ref{f1} it
is apparent that the X-ray emitting electrons have extremely short
lifetimes ($\sim$ 20 years). This fact, coupled with the known 
projected size of
the PWN in X-rays ($\sim$0.3~pc), implies that the propagation speed of
electrons down-stream from the pulsar should be $\sim$10\% of the
speed of light, consistent with expectations for PWN (see for example
\citet{blondin} and \citet{kennel}).

It is clear from Fig.~\ref{f4} that the shape of the H.E.S.S.  and
Chandra spectra can be explained in broad terms in this scenario.
G\,359.95$-$0.04 may produce $\sim100$ GeV $\gamma$-rays very
efficiently despite its high $B$-field.  Fine-tuning of the model and
adjustment of radiation fields would be required to fit all
H.E.S.S. spectral data points but we consider such tuning unjustified
given the possibility of the contribution of other sources (the
supernova remnant Sgr~A East, Sgr~A$^{\star}$, etc) to the
H.E.S.S. signal, and since there are larger systematic errors on the
H.E.S.S. spectrum close to threshold. It therefore appears that
G\,359.95-0.04 is a promising counterpart to the TeV GC source. As can
be seen from Fig.~\ref{f2}, a PWN like G\,359.95$-$0.04 could not be
detected by any current or planned $\gamma$-ray detector if located in
a standard region of the galactic disc. The existence of extended
regions with comparable radiation densities outside of the GC, for
example in the central regions of young stellar clusters, seems
unlikely since $\sim1000$ O-stars would be required within 1 cubic
parsec to reach this energy density.

\subsection{Comparison with G\,0.9+0.1}

G\,0.9+0.1 is a composite SNR with a bright radio shell and a compact
core~\citep{helfand}. The central object was identified as a PWN
based on its X-ray properties~\citep{mereghetti,xmm_g09}. VHE $\gamma$-ray
emission associated with the PWN has been reported by the
H.E.S.S. collaboration \citep{hess_g09}, with energy flux comparable
to that emitted in X-rays ($\sim3\,\times\,10^{-12}$ cm$^{-2}$
s$^{-1}$).  The PWN in G\,0.9+0.1 can be considered an intermediate
case between a `standard' disk PWN and G\,359.95-0.04. With a
projected distance from GC of 100~pc, radiation fields close to
G\,0.9+0.1 are likely a factor 10--100 higher than local densities, as
represented by radiation field R2 in Table~\ref{t1} 
\citep[see for example][]{moskalenko}.  Indeed an increased energy density of
8~eV\,cm$^{-3}$ in optical photons was invoked in \citet{hess_g09} where
a time-independent IC model\footnote{i.e. a model for the
present-day electron spectrum without consideration of the injection
spectrum required to produce such a spectrum after cooling} was fit to the
X-ray and $\gamma$-ray data. Here we revisit the SED of G\,0.9+0.1
from the perspective of a time-dependent evolution of the electron
spectrum.  In the following we assume that G\,0.9+0.1 is located at
the same distance as the GC (7.6 kpc).

\begin{figure}[ht]
\plotone{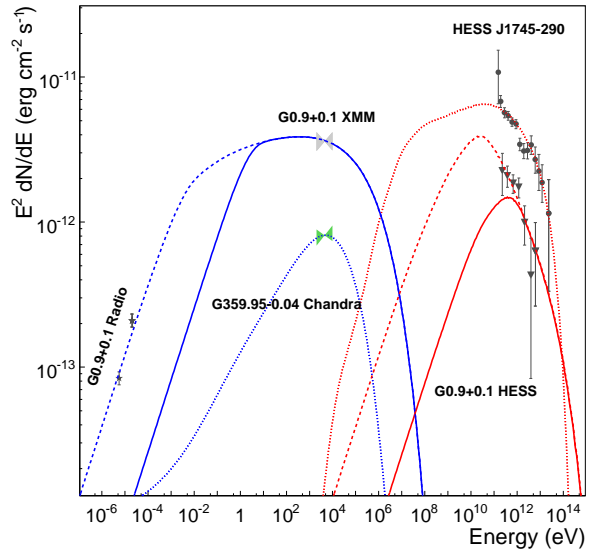}
\caption{
  Comparison of the G\,359.95-0.04 SED shown in 
  Fig.~\ref{f4} (dotted model curves) 
  with G\,0.9+0.1 (solid and dashed curves). The solid and
  dashed curves differ only the assumed age of the source.
  Model parameters are given in the main text. 
  Radio data are taken from \citet{helfand}, XMM data from
  \citet{xmm_g09}.
}
\label{f5}
\end{figure}

Using values of the optical and FIR densities of 9 eV\,cm$^{-3}$ and 1
eV\,cm$^{-3}$ respectively, as suggested by \citep{moskalenko} for the
wider GC region, we find that a model with $\alpha=2$ , $L_{e} =
7\times10^{36}$ erg/s, $B=10$ $\mu$G and a pulsar age of 40~ky, is
consistent with the H.E.S.S. and X-ray data (see Fig.~\ref{f5}).  This
value is significantly larger than the 6800 year age of the remnant estimated assuming
expansion in the Sedov phase~\citep{mereghetti}.
Pulsar ages much shorter than 40~ky are excluded by the 
absence of a spectral maximum (produced by a cooling break in the 
electron spectrum) within the H.E.S.S. energy range. 
Recently, \citet{porter} have fit the 'prompt' electron 
spectrum of G\,0.9+0.1 using the radiation field of \citet{moskalenko}
and derive a magnetic field of $B=9.5$~$\mu$G very close
(as expected) to the value given here. 

A second component of lower energy electrons is required to explain
the radio emission~\citep{helfand} unless an age of $\approx 6
\times 10^{5}$ years is assumed (see dashed lines in
Fig.~\ref{f5}). \citet{sidoli} used the break energy implied by the
combination of radio and X-ray data to estimate the age of the pulsar
to be $\sim3000$ years (assuming $B=67$~$\mu$G). The much larger age
derived here is a consequence of the lower magnetic field
value established by the combination of X-ray and $\gamma$-ray data.

\section{A central 10 parsec source}

\begin{figure}[ht]
\plotone{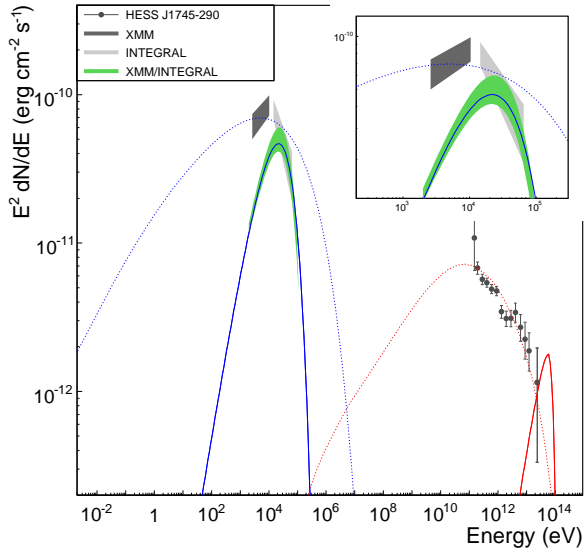}
\caption{ Spectral energy distribution for a central 10~pc
  XMM/INTEGRAL/H.E.S.S. source. Separate XMM and INTEGRAL data are
  taken from \citep{neronov}, the combined XMM/INTEGRAL fit is that of
  \citet{belanger}. Model curves are shown for two scenarios: a very
  young source with $B=50$~$\mu$G, electron spectrum with $\alpha = 0.3$
  and a sharp cut-off at 100~TeV (solid curve), an old source with $B
  = 110$ $\mu$G and an injection spectrum with $\alpha=1.5$ and an
  exponential cut-off at 150~TeV (dashed curve). The inset panel
  provides an expanded view of the X-ray part of the SED.
  \label{f6}
}
\end{figure}

The recently detected hard (20--100 keV) X-ray source
IGR\,J1745.6$-$2901~\citep{belanger04}, is located within $1'$ of
Sgr~A$^{\star}$ and coincident with HESS\,J1745$-$290.  The angular
resolution of INTEGRAL (12$'$ FWHM) is comparable with that of
H.E.S.S. and similar difficulties exist with the identification of a
counterpart. However, a combination of the INTEGRAL data with that of
XMM suggests that the INTEGRAL source represents the sum of the
emission of the central $\sim$20 parsecs, either from a diffuse
component or the combination of several discrete sources~
\citep{neronov,belanger}. The combined XMM/INTEGRAL X-ray spectrum of
the central region has been derived by both groups, 
but with somewhat different results.  \citet{neronov}
provide a broken power-law fit with
$\Gamma_{1}=1.85^{+0.02}_{-0.06}$ and
$\Gamma_{2}=3.3\pm0.1$ with a break at $26\pm1$ keV (hereafter the BPL fit).
\citet{belanger} provide a broken power-law fit:
$\Gamma_{1}=1.51^{+0.06}_{-0.09}$ and
$\Gamma_{2}=3.22^{+0.34}_{-0.30}$ with a break at
$27.1^{+2.8}_{-4.4}$ keV and also a cut-off power law fit: $\Gamma
=1.09^{+0.03}_{-0.05}$, $E_{\mathrm{cut}} = 24.38^{+0.55}_{-0.76}$
keV (PLEC).  Of these three fits, BPL and PLEC represent the 
extremes and are used here to illustrate the impact of the X-ray 
spectral shape on the interpretation.
If this emission has a synchrotron origin, then the
broken power-law fit suggests a change in the electron spectral slope
of $\sim3$, incompatible with the effects of cooling or escape, and with
standard acceleration scenarios. It therefore seems that the INTEGRAL
data represent the end of the X-ray spectrum of this object or
objects. Fig.~\ref{f6} shows approximate error boxes corresponding to
the spectral fits BPL and PLEC. The difference in low energy
slope of these two fits has important implications for the
interpretation of this signal in a synchrotron scenario. In either
case an abrupt end to the electron spectrum is required to produce an
exponential cut-off in the synchrotron spectrum. The PLEC fit implies
an extremely hard electron spectrum.

Although considerable uncertainty exists in the radiation density
$\sim$10~pc from the GC, it seems likely that the average energy
density within the 40~pc diameter INTEGRAL source is roughly two
orders of magnitude lower than that within the central parsec \citep[see for
example][]{yusef_zadeh}. For the calculation of the inverse
Compton emission we therefore assume the radiation field R4 given in
Table~\ref{t1}. With this radiation density, the association of
HESS\,J1745-290 with IGR\,J1745.6-2901 implies a $B$-field of $\sim100$
$\mu$G. In the case that these objects are not associated this $B$-field
can be considered as lower limit on the mean value within the source
or sources contributing to the INTEGRAL signal.  Conversely if the
$\gamma$-ray emission has an inverse Compton origin within a 10~pc
scale source then the $B$-field in this region must be $< 100$~$\mu$G 
to avoid over-producing synchrotron emission.  For a 100~$\mu$G
magnetic field, 20~keV synchrotron photons are produced by $\sim70$~TeV
electrons. The cooling time of these electrons in such a field is
extremely short ($\sim 18$ years). The time required for electrons to
diffuse out of the central 10~parsecs is comparable ($\sim$10 years)
if a diffusion coefficient close to that appropriate for 70~TeV galactic
cosmic rays is assumed (i.e. $D \approx 6 \times 10^{30}$ cm$^{2}$
s$^{-1}$). It therefore seems rather difficult to produce a truly
diffuse 20~pc source in the presence of such rapid energy losses. This cooling-time
problem was previously discussed by \citet{neronov}. A model similar to
that of \citet{loeb}, with acceleration occurring at
stellar wind shocks, could avoid this problem by distributing the
acceleration sites of electrons over the emission region.

Fig.~\ref{f6} shows two model curves illustrating the relationship
between the X-ray and $\gamma$-ray emission. The first (solid) line
matches the PLEC fit to the X-ray spectrum but can explain only the
highest energy H.E.S.S. points. This scenario requires both an
extremely young source (to avoid a cooled spectrum in the XMM domain)
and a very hard injection spectrum. The second curve (dashed line) is
similar to that given in \citet{neronov} and provides marginal
agreement to the BPL fit and reasonable agreement with the
H.E.S.S. spectral data. The dramatic difference between these two 
scenarios illustrates the importance of better constraints on the
X-ray spectrum and highlights the value of combined X-ray/$\gamma$-ray
measurements.

\section{Summary}

The central $\sim$10~pc of our galaxy provides a unique environment in
which high radiation energy densities lead to efficient inverse
Compton $\gamma$-ray production, and also, due to the Klein-Nishina
effect, to substantial modifications to the form of cooled electron spectra. This region appears to be the
only location in our galaxy in which pulsar wind nebulae with high
magnetic fields and moderate spin-down luminosities can produce
detectable $\gamma$-ray emission.  In this context, the candidate PWN
G\,359.95$-$0.04 provides a plausible counterpart to the $\gamma$-ray
source HESS\,J1745$-$290.  The interpretation of IGR\,J1745.6$-$2901
and HESS\,J1745$-$290 in terms of synchrotron/IC emission in a diffuse
20~parsec source is difficult due to the rapid energy losses of
electrons in the region, but remains a viable alternative hypothesis.
Finally, hadronic models for the TeV emission, while beyond the scope
of this paper, provide equally viable explanations for the current
experimental data.
Improved $\gamma$-ray data should be available in the medium term
from the GLAST~\citep{glast} and H.E.S.S. Phase-2~\citep{hess2} 
instruments and will provide important constraints on the origin 
of the high energy emission.

\acknowledgments

JAH acknowledges the support of the German BMBF 
through Verbundforschung Astro-Teilchenphysik (05CH5VH1/0).
We are grateful to Farhad Yusef-Zadeh for providing a radio upper
limit for G\,359.95-0.04. We would also 
like to thank Mitya Khangulyan for many useful 
discussions and Karl Kosack for his careful reading of the manuscript.

\end{document}